# Improving Rating and Relevance with Point-of-Interest Recommender System

Syed Raza Bashir
Department of Computer Science
*Ryerson University*
Toronto, Canada
syedraza.bashir@ryerson.ca

Vojislav Misic
Department of Computer Science
*Ryerson University*
Toronto, Canada
vmisic@ryerson.ca

*Abstract*— The recommendation of points of interest (POIs) is essential in location-based social networks. It makes it easier for users and locations to share information. Recently, researchers tend to recommend POIs by treating them as large-scale retrieval systems that require a large amount of training data representing query-item relevance. However, gathering user feedback in retrieval systems is an expensive task. Existing POI recommender systems make recommendations based on user and item (location) interactions solely. However, there are numerous sources of feedback to consider. For example, when the user visits a POI, what is the POI is about and such. Integrating all these different types of feedback is essential when developing a POI recommender. In this paper, we propose using user and item information and auxiliary information to improve the recommendation modelling in a retrieval system. We develop a deep neural network architecture to model query-item relevance in the presence of both collaborative and content information. We also improve the quality of the learned representations of queries and items by including the contextual information from the user feedback data. The application of these learned representations to a large-scale dataset resulted in significant improvements.

## I. INTRODUCTION

Users of location-based social networks like Foursquare and Yelp can share their check-in experiences and views on the POIs that they've visited. A POI is a specific point location that someone may find useful or interesting. POI recommendation [1], as one of the key services of Location-based Social Networks (LBSNs), focuses on mining user check-ins, location-based information, and social ties to recommend POIs to users. The POI suggestion helps improve user experiences, social networking services, and advertising between prospective locations and users. Many methods [2]–[5] are proposed in the literature to solve POI recommendations. The seminal work in this regard focuses mainly on utilizing the collaborative and content information to provide POI recommendations and retrieve the top-k relevant items. However, a few more challenges are specific to developing a real-world, large-scale POI recommendation system, which are discussed next.

*Sparse feedback data*: the user-location matrix in a POI recommender is primarily sparse. This sparsity is mainly because users in a POI recommender visit a few locations and do not explicitly share their experiences. The sparsity issue also happens when an individual travels to a new place, and the POI recommender does not have enough information to provide to the user. Some works [6], address this sparsity issue by using human-evaluation templates such as Likert scale questions and request feedback using crowd-sourcing platforms (e.g., Amazon Mechanical Turk), which is both expensive and nearly impossible when it comes to Big Data.

*Incomplete user-item information*: Usually, these POI recommenders [7] are built using the collaborative filtering (CF) methods. These traditional CF-based recommender systems typically do not consider the contextual attributes of users or the meta-data information related to items while making recommendations. The 'context' refers to a situation in which a user is currently in [8]. For example, in a POI recommender, the context can be the location and time during which the user has interacted with the system. In a POI recommendation scenario, the choices made by a person may not only depend on his or her earlier preferences but more on the context. Therefore, we need to exploit the contextual information involved with POI check-ins (i.e., POI category, location, or the check-in timestamp) and provide recommendations to a user that are tailored to his personal preferences.

In this work, we propose to address the sparse feedback and incomplete user-item matrix using a deep learning approach. Specifically, we propose a novel deep learning model for large-scale POI retrieval systems. We develop a deep learning model that learns the effective user and item representations using users' feedbacks (ratings) from a POI data. Our model can predict those POIs that a user visited and which they did not. To accomplish the recommendation task, we build two sub neural network models: a user encoder that turns user-related information into user embeddings (high-dimensional vector representations); and a location encoder that turns business (location) related information into location embeddings We define a scoring function that maximizes the predicted user and location affinity for the visited locations and minimizes it for visits that did not happen. The aim here is to train the proposed model using the user and item encoder), such that POI items with embeddings very similar to the user are recommendations for the user.

Our proposed deep neural network is based on the multi-task learning [9]. Multi-task learning is a machine learning approach that learns multiple tasks simultaneously while optimizing multiple loss functions. We compute the loss for two different tasks in one model, which are:
1) Rating task: we treat the rating task as a regression problem and fit the mean squared error (MSE) loss function on users' feedbacks.
2) Retrieval loss: we facilitate efficient retrieval of candidates and use the cross-entropy loss function on users' feedbacks.

Our model also considers the auxiliary information (related to the location item) and contextual information (related to the user) in each tower. The intuition is that we can learn high-quality representations from the data if we consider additional information to build a POI recommendation model.

We list our contributions as:
1) We propose a deep learning framework to learn better representations to address the data sparsity problem.
2) We build a user encoder and a location encoder that compute user and item embeddings. We include the auxiliary information from the location encoder and the user's contexts from the user encoder to build a strong predictive and ranking location-aware recommendation model.
3) We design a novel model based on multitasking that optimizes two training objectives: optimizing the rating and retrieval tasks.

We test our model on a real-world benchmark dataset and find that it improves significantly.

The rest of the paper is organized as follows: Section II describes related work. Section III presents the proposed methodology and Section IV describes the experimental setup. Section V discusses the results and analysis. Finally, Section VI concludes the paper.

## II. RELATED WORK

This section covers some state-of-the-art POI recommender systems, as well as the implementation of multi-task learning approaches in POI retrieval and recommendation tasks

Retrieval systems are commonly employed in large-scale applications like POI recommendations [10]. In the past few years, the industry has shifted away from reverse index-based retrieval systems [11] and toward machine-learning retrieval systems [12]. Collaborative-filtering [13] based systems have been very popular and successful until very recently, when they were surpassed by various neural network based retrieval models [3], [14], [15].

Two main components of a retrieval system are representation learning (feature engineering) [16] and efficient indexing techniques [17]. Using different representations for query and candidate items has shown to be successful in a large-scale POI retrieval system too [18]. There has also been work on multi-task learning for POI applications [19]. However, the model cannot be easily extended to learn complicated feature representations from the large POI data.

A content-based filtering POI model [20] exploited visitors' explicit ratings to recommend tours in a museum. The other widely exploited technique used in POI recommender systems is CF technique [21]. The above-mentioned existing studies have exploited few contexts and have focused on personalized POI visits. Unlike others, our work exploits contextual information (temporal, spatial) to generate the POI recommendations. We also exploit the auxiliary information related to the POI items (reviews on items, check-in, category) to make recommendations. In comparison to the existing studies, our paper also combines two important tasks: (i) rating specialized and (ii) retrieval specialized in the same unified model. Our model can learn from rich feedback information to help with data sparsity problem in a POI recommender system.

## III. METHODOLOGY

We formalize the problem and introduce our training data and training objectives in this section.

### A. Problem Definition

We treat the problem of our proposed POI recommender system as a query - candidate model [22]. Our model uses the similarities between queries and items simultaneously to provide recommendations. For example, if user A is similar to user B, and user B likes item 1 (POI) then the system can recommend item 1 to user A (even if user A hasn't visited any places similar to item 1). Precisely, the problem can be defined as:

Given a query and a corpus of candidate POI items, return the top-k relevant items.

Let $\{x_i\}_{i=1}^{N} \subset \mathcal{X}$ denote the feature vectors of queries ($N$ is the number of queries) and $\{y_j\}_{j=1}^{M} \subset \mathcal{Y}$ represents the features vectors of candidates ($M$ represents the number of candidates). The outputs from the query representations and the candidate representations are then multiplied together to give a query-candidate affinity score, with higher scores expressing a better match between the candidate and the query. The scoring function for the recommendation model is then defined as shown in Equation (1):

$$s(\cdot,\cdot;\boldsymbol{\theta}): \mathcal{X} \times \mathcal{Y} \mapsto \mathbb{R} \qquad (1)$$

where $\boldsymbol{\theta}$ denotes the model parameters. The items with top-k scores $s(x,y;\boldsymbol{\theta})$ are selected for a given query.

In this work, we consider the training data as a set of query and item pair $\{(x_t, y_t)\}_{t=1}^{T}$, where $T$ is the number of training examples, $y_t$ is the candidate associated with $x_t$ that has either explicit or implicit users' feedback, and $T \ll MN$.

Our goal is to fit the scoring function based on these $T$ examples. In this paper, we study the retrieval systems with explicit feedback (1-5 star ratings), however, the same model can be used to include implicit feedback.

### B. Optimization

In this optimization problem, each query example $(x_t, y_t)) \in \varepsilon$ is associated with a label $l_t \in \mathbb{R}$ representing user' feedback. Our training objective is to minimize the differences between the predicted relevance and ground truth.

In this work, we propose a multi-tasking learning where we propose two specific tasks i.e., a *rating-specialized* task and a *retrieval-specialized* task. We have a separate loss function for each of task: $\mathcal{L}_{rating}$ for rating and $\mathcal{L}_{relevance}$ for relevance task. We give each of the losses a weight and treat these weights as hyperparameters. If we assign a large loss weight to the rating task, our model will focus on predicting ratings (but still use some information from the retrieval task); if we assign a large loss weight to the retrieval task, it will focus on retrieval.

Formally, we define these two tasks as:

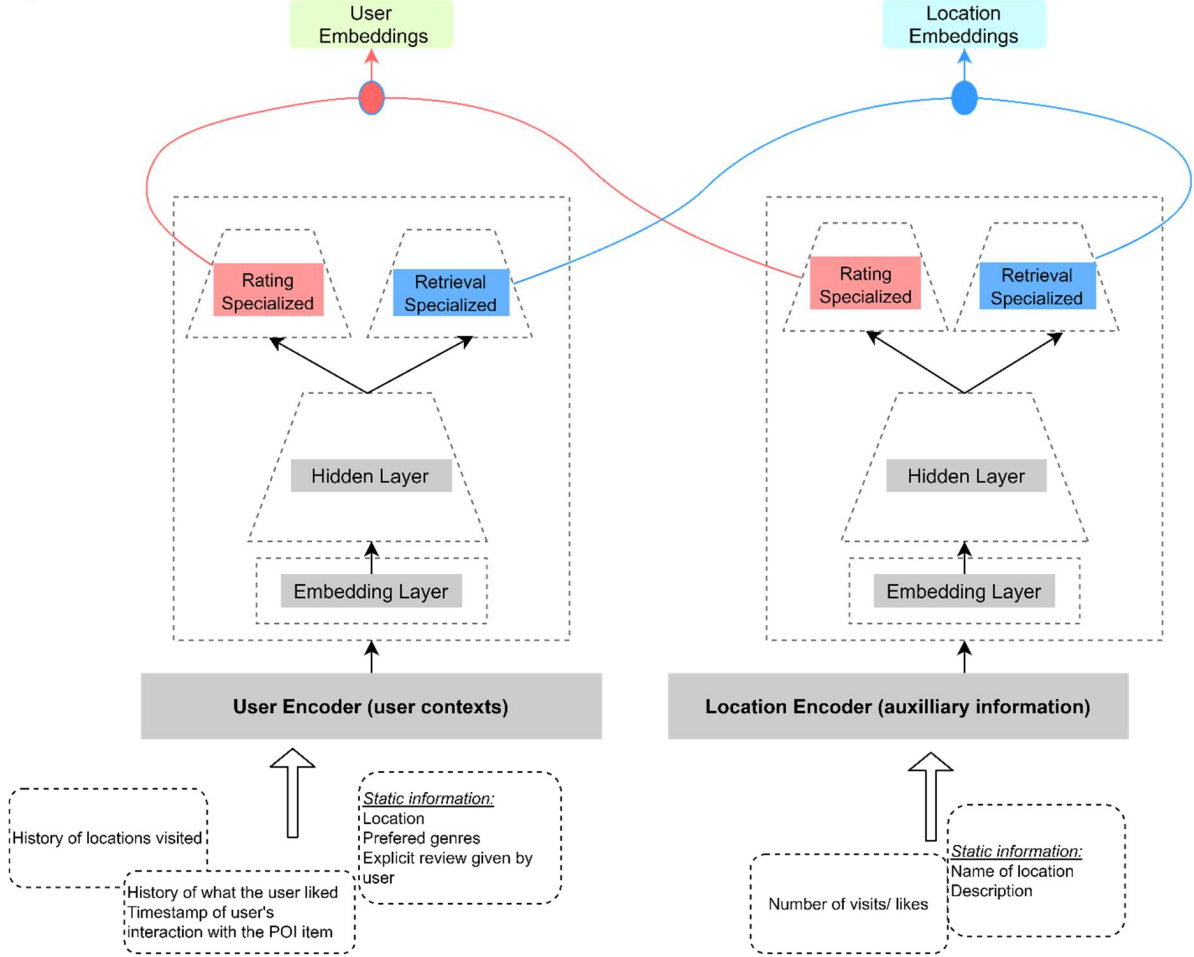

Fig. 1: Proposed Model Architecture

*Rating-specialized task*

In this task, our model is going to focus on predicting ratings. We use regression loss to fit users' feedback on the example set in $\varepsilon$. This loss function is the mean squared error (MSE), which is shown in Equation (2):

$$\mathcal{L}_{rating}(\boldsymbol{\theta}; \varepsilon) = \frac{1}{|\varepsilon|} \sum_{(x_t, y_t) \in \varepsilon} \left(s(x_t, y_t; \boldsymbol{\theta}) - l_t\right)^2 \quad (2)$$

Where $\mathcal{L}_{rating}$ is rating loss function, $t$ is training example, and $|\cdot|$ represents the cardinality of examples.

*Retrieval-specialized task*

This task facilitates efficient retrieval of candidates from a large POI corpus. In this work, we treat the modeling of feedback as a multi-class classification task over the full corpus of items, and use the Categorical Cross-Entropy loss [23] formulation to model the probability $\mathcal{P}$ of choosing item $y$, as shown in Equation (3):

$$\mathcal{P}(y \mid x; \boldsymbol{\theta}') = \frac{exp(s(x, y; \boldsymbol{\theta}'))}{\sum_{j=1}^{M} exp(s(x, y_j; \boldsymbol{\theta}'))} \quad (3)$$

Where $\boldsymbol{\theta}'$ denotes the model parameters with retrieval task. The maximum likelihood estimation (MLE) is formulated as shown in Equation (4):

$$\mathcal{L}_{retrieval}(\boldsymbol{\theta}', \varepsilon) = -\frac{1}{|\varepsilon|} \sum_{(x_t, y_t) \in \varepsilon} \log \mathcal{P}(y_t \mid x_t; \boldsymbol{\theta}') \quad (4)$$

*Joint optimization*

Formally, we aim to jointly learn two objectives $\mathcal{L}_{rating}(\cdot, \cdot; \boldsymbol{\theta})$ and $\mathcal{L}_{retrieval}(\cdot, \cdot; \boldsymbol{\theta}')$ while sharing part of the parameters between $\boldsymbol{\theta}$ and $\boldsymbol{\theta}'$. With loss multipliers $w$ and $w'$, we jointly optimize the losses in Equations (2) and (4) by optimizing, as shown in Equation (5):

$$\mathcal{L}(\boldsymbol{\theta}, \boldsymbol{\theta}') = w \cdot \mathcal{L}_{rating}(\boldsymbol{\theta}; \varepsilon) + w' \cdot \mathcal{L}_{retrieval}(\boldsymbol{\theta}'; \varepsilon) \quad (5)$$

### C. User-Location Encoder Architecture

Our proposed model also consists of two main parts:

The **user encoder,** which is a model responsible for learning the user embeddings. The user encoder learns the history of the POI items the user liked and time stamps of user's interaction with the items. The user encoder can also contain other metadata of the user if shared like genres (restaurants, hotels, parks, museums, etc.) preferred, or any past interactions. In a social context like Twitter, one could also potentially consider attributes of other users that this user is following / connected to.

The ***location encoder***, which is a model responsible for learning the POI embeddings. It may also consider the title, description of the POI item. Other metadata like the language, number of visits can also be considered.

Our user-item encoder model is shown in Figure 1.

Given a pair of query and item (POI) represented by feature vectors $x \in \mathcal{X}$, $x \in \mathcal{Y}$, respectively, we present the embedding functions $u : \mathcal{X} \times \mathbb{R}^d \mapsto \mathbb{R}^k$, $v : \mathcal{Y} \times \mathbb{R}^d \mapsto \mathbb{R}^k$, which encode features of user and item to a *k*-dimensional embedding space. The scoring function is then computed as the dot product between the query and item embeddings at the top layer, as shown in Equation (6):

$$s(x, y; \boldsymbol{\theta}) = \langle u(x, \boldsymbol{\theta}), v(y, \boldsymbol{\theta}) \rangle \quad (6)$$

We have extended the proposed model (shown in Figure 1) to enable multi-task learning (Equation 5). Specifically, we introduce two hidden layers, one for the rating-specialized task and the other for the retrieval-specialized task. The outputs of the hidden layers are fed in parallel to the proposed model. The multi-task learning optimizes both the objectives (rating and relevance) for both loss functions (Equations 2 and 4).

## IV. EXPERIMENTAL SETUP

### A. Data Set

In this paper, we used the Yelp Dataset[i], which can be found on Yelp website. The dataset contains 1.6M reviews and 500K tips by 366K users for 61K businesses; 481K business attributes, such as hours, parking availability, ambience; and check-ins for each of the 61K businesses were gathered over time via a social network of 366K people for a total of 2.9M social edges. Precisely, it includes the data of 61184 businesses, 1569264 reviews and 366715 users. We used the following information from the dataset, as shown below:

{ 'type': 'review',
  'business id': (encrypted business or location id),
  'user id': (encrypted user id),
  'stars': (1-5 star rating),
  'text': (review text),
  'date': (date, formatted like '2016-05-30'),
  'votes': (vote type: funny, useful, cool): (count) }

We perform some exploratory analysis on the Yelp dataset and explore the distribution of different stars (1–5-star rating) and the lengths of the review texts corresponding to those ratings, as shown in Fig. 2.

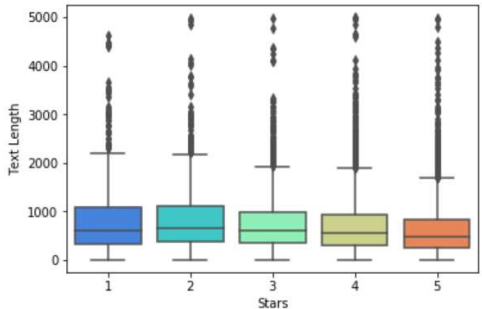

Fig. 2: Stars and Text (review) Lengths

### B. Evaluation Metric

We test our model using two evaluation metrics:

*Root Mean Square Error* (RMSE) measures the difference between values predicted by a model and the actual values.

*top-K categorical accuracy metric*[ii], which calculates the percentage of records for which the targets (actual values) are in the top-K predictions. This metric divides the number of accurately predicted records by the total number of records.

We also use precision, recall, F1-score in a confusion matrix in one of the experiments.

### C. Hyperparameter Optimization

We temporally split the data into 90% for training and 10% for evaluation. The model was implemented in Tensorflow using the GPUs provided by Google Colab. We train our model using Adagrad optimizer [24] with a learning rate of 1e-3 and epsilon value of 1e-07. Each experiment is repeated 10 times. All the other models (baselines) are also trained from scratch. For fair comparison, the hyper-parameters including model size, learning rate, and training steps were carefully tuned for the best model performance of each baseline.

### D. Baselines

We use the following baselines:

*Neural Collaborative Filtering* (NCF) [25] that is a multi-layer neural network to perform collaborative filtering.

*Matrix Factorization* (MF) [26] is considered a de-facto model for the collaborative filtering recommendation tasks. We use the MF model with temporal information.

*Deep Matrix Factorization* (DMF) [27] that is matrix factorization model based on neural network architecture.

*BERT4REC* [28] is a sequential rrecommender system with Bidirectional Encoder Representations from Transformer.

### E. Training

We train the model for the rating-specialized task during training, which uses the MSE loss. After learning the shared representations, we finetune the model for the retrieval-specialized objective that uses the cross-entropy loss. To prevent potential over-fitting, we apply stop gradients on the layers.

## V. RESULTS AND ANALYSIS

### A. Overall Model Performance

Our model comparison with the baselines is shown as below in Fig. 3.

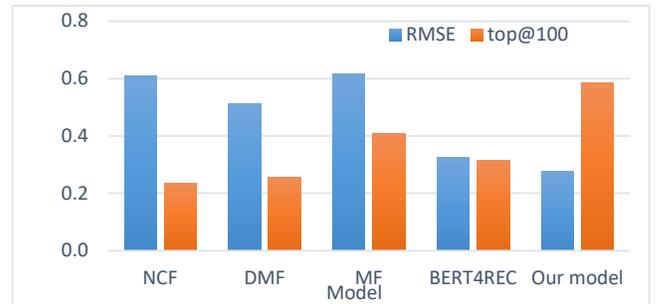

Fig. 3: Comparison with Baselines

We see in Fig. 3 that our proposed model outperforms all the baselines in terms of accuracy (lowest RMSE) and top@100 categorical accuracy. The superiority of our model is attributed to the multiple loss functions that we try to optimize. The other baselines (NCF, MF DMF and BERT4REC) optimize a single utility function. For example, the MF uses a loss function based on regularized squared error, NCF uses MSE and DMF optimizes cross-entropy loss. BERT4REC optimizes the negative log-likelihood. In contrast, we optimize the two losses (MSE and cross-entropy loss) using multi-tasking learning.

We also find from the results in Fig. 3 that the general performance of the sequential transformer-based model (BERT4REC) is better than CF models (NCF and DMF). The exception is with the MF model that has higher RMSE (lower accuracy) than BERT4REC, but it shows a better top@ k accuracy than the other baselines. The CF methods like MF have also shown good performance in the recommendation accuracy tasks, as shown in past work [26], [29].

The BERT4REC is based on the Transformer [30] model that can differentially weigh the significance of each part of the input data. Since BERT4REC is based on a sequential modelling paradigm, it has shown better performance on the Yelp dataset where we try to capture the time-related user-item interactions.

The DMF and NCF can also model the time-related user-item information, these models are usually limited to capture the dynamic information in user preferences [28]. The user-item interactions are usually dynamic in a recommender system. Our proposed model is not based on the self-attention layer as in the Transformer model but building a joint model for multiple tasks (rating and retrieval) may have produced better results than building several task-specific models (as in these baselines).

By using rich auxiliary and contextual information, our model is able to leverage additional information available from the data for recommendations. This facilitates us to mitigate the data sparsity issue that is quite prevalent in content-based and CF systems.

### B. Task-specific Performance

We also conduct an experiment to determine the importance of each loss in our model. This part of the experiment is carried out by assigning a weight to each of the losses and treating these weights as hyperparameters. If we give the rating task a large weight, our model focuses on predicting ratings. If we give the retrieval task a large loss weight, it will focus on retrieval instead. A weight of 0.5 means we are assigning equal weight to both the tasks. The results are shown in Table I.

TABLE I: MODEL VARIANTS

| Model | Rating weight | Retrieval weight | RMSE | Top@100 |
|---|---|---|---|---|
| Rating-specialized | 1.0 | 0.0 | 0.113 | 0.085 |
| Retrieval-specialized | 1.0 | 0.0 | 0.703 | 0.235 |
| Joint | 0.5 | 0.5 | **0.278** | **0.587** |

As shown in Table I, the rating-specialized model performs good on predicting ratings (with an RMSE of around 0.113) but performs poorly at predicting which POI the user will like. The top@100 categorical accuracy is quite poor because the loss function for the retrieval task is not given a weight in the rating-specialized model. For the retrieval-specialized model, we get the opposite result: the model performs well on retrieval (0.235) but poorly predicts ratings (0.703).

We also see that the joint model (with both loss functions) performs well on both tasks as each specialized model. Overall, the results show that the multi-task learning, in general, is an extremely useful tool. We can expect better results when we optimize the model for both tasks.

### C. Effectivness of Auxilliary Information and Contexts

We test the effectiveness of including multiple features into our joint model (joint model showed the best performance in Table I, to check whether those features improve model accuracy. In this experiment, we test the performance of model by including and excluding the auxiliary information (text – review) and the contextual information (date of user's interaction with the POI)

The results are shown in Table II, where we refer to our joint model as *M* and write the features inside the bracket. The user ID and the business ID are the main features. The auxiliary information is the text -review on the POI item and the date is the context when the user gives a rating to the POI.

TABLE II: MODEL VARIANTS

| Model with features | RMSE | Top@100 |
|---|---|---|
| M (user ID, business ID, rating) | 0.413 | 0.209 |
| M (user ID, business ID, text, rating) | 0.328 | 0.337 |
| M (user ID, business ID, text, date, rating) | 0.278 | 0.587 |

Overall, the results in Table II show that our joint model with more features give better results, compared to those model variants where one or two features are deducted. If the users' preferences are contextual as in a POI recommender system, adding auxiliary information will improve the model significantly. For example, the month may be an important feature when deciding whether to recommend a POI item: users may only have time to visit beaches during the summer months. Similarly, if we get to know the text of the review, we can better understand the preference of the user towards a POI item. Although, our 5-star rating are used as proxy to the textual review, but the research [31] shows that review texts contain richer information about users and items which improves the recommendation accuracy to some extent, that's why we get better accuracy when we include the text in the model.

In general, we find that auxiliary features (other than user and item ID) such as item categories, descriptions, and images must be used to help the model generalize and to build an accurate model. This is especially important in cold-start situations, where there is little data on some items or users.

We also perform an experiment to check the accuracy of the stars rating with respect to the text of the reviews. We simply convert those texts to vectors and use the Multinomial Naive Bayes (MNB) classifier [32]. MNB is a well-known model for text data analysis and can be used to test the accuracy of the text with multiple classes. The results are shown in a confusion matrix in Table III.

TABLE III: CONFUSION MATRIX

|  | Stars | precision | recall | f1 | support |
|---|---|---|---|---|---|
|  | 1 | 0.76 | 0.51 | 0.61 | 69 |
|  | 2 | 0.56 | 0.10 | 0.17 | 92 |
|  | 3 | 0.25 | 0.13 | 0.17 | 137 |
|  | 4 | 0.46 | 0.68 | 0.55 | 373 |
|  | 5 | 0.60 | 0.56 | 0.58 | 329 |
| micro | average | 0.50 | 0.50 | 0.50 | 1000 |
| macro | average | 0.53 | 0.40 | 0.42 | 1000 |
| weighted | average | 0.51 | 0.50 | 0.48 | 1000 |

The precision is a measure of result relevancy, while recall is a measure of how many truly relevant results are returned. F1-

score is a harmonic mean of precision and recall. Other details of confusion matrix can be found here[iii]. Overall, the results in Table III shows the correct and incorrect predictions with count values and broken down by each class (1–5-star rating) and we find better accuracies with 1 and 5 stars.

## VI. CONCLUSION

We propose a novel model architecture to learn better query and candidate representations via transfer learning. We leverage the auxiliary and contextual information with rich feedback data. By introducing multiple objectives and jointly learning this model via multitasking, we observe a significant improvement for relevance prediction tasks on a POI retrieval systems.

In future, we plan to include more layers in our proposed model. For example, we would include the self-attention layer to see if it further improves the system. We will also include more feedback information (implicit as well as explicit) from the POI data. We would also like to use representations learned from the abundant task to improve its predictions on the sparse task via a phenomenon known as transfer learning [14]. We like to consider different weights to assign to each of the task (retrieval and rating) to check that how it affects the performance. Finally, we want to test our model against more robust baselines and using multiple evaluation metrics.

---

[i] https://www.yelp.com/dataset
[ii] TopKCategoricalAccuracy
[iii] Confusion Matrix